\documentclass[aps,twocolumn,prb,floatfix,showpacs]{revtex4}
\usepackage{graphicx}
\usepackage{longtable}
\usepackage{color}

\begin{document}
\pagestyle{plain}

\title{Selective Heating Mechanism of Magnetic Metal Oxides 
by Alternating Magnetic Field in Microwave Sintering Process}

\author{Motohiko Tanaka$^{1}$, Hirohiko Kono$^{2}$, and
Koji Maruyama$^{3}$}

\affiliation{$^{1}$Coordination Research Center, National Institute 
for Fusion Science, Toki 509-5292, Japan}
\affiliation{$^{2}$Graduate School of Science, Tohoku University, 
Sendai 980-8578, Japan}
\affiliation{$^{3}$Advanced Science Institute, RIKEN, 
Wako 351-0198, Japan}

\begin{abstract}
The mechanism of rapid and selective heating of magnetic 
metal oxides under the magnetic field of microwaves 
which continues beyond the Curie temperature $ T_{c} $  
is identified by using the Heisenberg model.  
Monte Carlo calculations based on the energy principle
show that such heating is caused by non-resonant response 
of electron spins in the unfilled 3d shell to the wave 
magnetic field. 
Small spin reorientation thus generated leads to a large 
internal energy change through the exchange interactions
between spins, which becomes maximal around $ T_{c} $ 
for magnetite $ {\rm Fe}_{3}{\rm O}_{4} $. 
The dissipative spin dynamics simulation yields 
the imaginary part of the magnetic susceptibility, 
which becomes largest around $ T_{c} $ and for the 
microwave frequency around 2 GHz.
Hematite $ {\rm Fe}_{2}{\rm O}_{3} $ with weak spontaneous 
magnetization responds much less to microwaves as observed 
in experiments. 
The heating of titanium oxide by microwave magnetic field 
only when oxygen defects are present is also explained 
by our theory in terms of the absence of spontaneous 
magnetization.
\end{abstract}


\maketitle

\section{Introduction}

Microwave sintering is the process in which electromagnetic energy of 
microwaves is delivered directly to electrons or atoms in materials. 
It enables rapid heating that leads to melting and recrystallization 
of original solid matters. 
A laboratory experiment demonstrated that a grained magnetite
sample was heated to above $ 1300^{\circ}$C within a minute 
by applied microwaves\cite{Roy}. 
The amount of energy consumption can be reduced and the release 
of CO$_{2} $ gas be halved in the microwave iron making which
reduces energy and environmental problems in modern iron 
industry\cite{Nagata}. 
Later it was shown that various metallic oxides including magnetite 
and titanium oxide with oxygen defects TiO$_{2-x}$ ($ x >0 $) were 
sintered quickly at the magnetic field maximum (i.e. the electric 
field null) in the microwave cavity experiments\cite{Agrawal1}.
The product sintered in the microwave magnetic field was shown
to have good magnetic properties as magnets.

The sintering of magnetite/hematite powders of micron sizes 
is characterized by selective heating.  
Only the domain of magnetite $ {\rm Fe}_3{\rm O}_4$ with strong 
magnetization was heated to $ 1300^{\circ}$C in the microwave 
magnetic field\cite{Sato}, which is much above the Curie temperature 
$ T_{c} $ ($585^{\circ}$C). 
The adjacent domain of hematite ${\rm Fe}_2{\rm O}_3$ remained 
at low temperatures.
The microwave sintering of magnetic materials is a non-resonant 
process that occurs with or without a static magnetic field and 
for any amplitude of wave magnetic field at a fixed microwave 
frequency. 
With ferromagnetic metals, microwave resonance absorption\cite{FMR0} 
and the non-resonant bolometric effect on dc resistance\cite{FMR1,FMR2,FMR3} 
were observed in ferromagnetic resonance (FMR) experiments under microwaves.
These phenomena required a static magnetic field of specific strength
to have resonance with microwaves, and were observed mainly below room
temperature, which are attributed to Joule heating or eddy current.
The microwave sintering, which occurs under wide range of strength and
frequency of microwave magnetic field without a static magnetic field
is physically a different process from the FMR process.

To date, the heating of dielectric materials by the microwave
electric field was extensively studied\cite{Rybakov}. 
We examined the heating of water, salt-contained water and ice 
by microwaves using molecular dynamics simulation\cite{Tanaka1,Tanaka2}. 
The heating was attributed, respectively, to excitation of electric 
dipoles, the Joule heating of salt ions, and the weakening of the 
hydrogen-bonded H$_{2}$O network by the presence of salt ions. 
(Pure ice was not heated by 2.5GHz microwave due to tight hydrogen 
bonds of water molecules.)
For bulk metals, microwaves are reflected at the surface due to
skin effects and do not transfer energy, while they penetrate into 
grained metallic powders (10$\mu$m in diameter) for a few centimeters 
and heat them\cite{Suzuki}. 
However, the sintering mechanism of metal oxides that have spontaneous
magnetization under the magnetic field of microwaves has not 
been resolved.

In this paper, we show theoretically that the sintering of 
particles of magnetic metal oxides, including magnetite 
and titanium oxide with oxygen defects, by a microwave 
magnetic field is due to non-resonant response of electron spins 
in the unfilled 3d shell.
We first use the energy principle and perform the Monte Carlo
simulations.
A small spin perturbation in response to an alternating external
magnetic field results in a large internal energy change 
through exchange interactions.
Next, studies in the time domain are done by performing 
dissipative spin dynamics simulation and detecting the linear 
response of spins. 
The temperature and frequency dependence of the heating rate 
by the microwave magnetic field is obtained on the basis of the 
imaginary part of magnetic susceptibility $ \chi_{i} $.
These results agree well with those by the energy principle, 
and $ \chi_{i} $ quantitatively accounts for the 
rapid heating of magnetite in the sintering experiments.

\section{Numerical Procedures}

   We have used the following procedures in our numerical 
simulations. The magnetization of magnetite and hematite is
well described by the Heisenberg model above the Verwey transition 
temperature (120 K)\cite{Cullen} since electrons are roughly 
localized\cite{Kittel}. 
The internal energy $ U $ of the magnetic system is represented 
by the three-dimensional spin vector $ {\bf s}_{i}$ of the electron 
at the i-th site, the exchange interaction coefficient $ J_{ij}$
between the i-th and j-th sites, and the external magnetic field 
$ {\bf B}_{w} $ of microwaves, which reads
\begin{eqnarray}
\label{eq.1}
  U({\bf B}_{w})= - \sum_{i,j} J_{ij}{\bf s}_{i} \cdot {\bf s}_{j}
    + \sum_{i} g \mu_{B} {\bf s}_{i} \cdot {\bf B}_{w}.
\end{eqnarray}
The summation of the exchange interactions in the first term 
is taken over the pairs of nearest neighbor sites. 
The magnitude of the spin vector satisfies $ |{\bf s}_{i}| = S $
for the spin angular momentum S. 
The second term, which may be called the Zeeman term, 
is the scalar product of magnetization 
$ {\bf M}= - \sum_{i} g \mu_{B} {\bf s}_{i} $ and the magnetic field, 
$ - {\bf M} \cdot {\bf B}_{w} $, where $ g \cong 2$ and 
$ \mu_{B}= e \hbar/2mc $.
We note that the Weiss field (the internal magnetic field) 
is $ B_{Weiss} \cong
n_{B} J_{AB}S/g\mu_{B} \cong 240 $ T, which is much larger than 
that of microwaves, where $ n_{B} \cong 4.1 $ is the effective 
magneton number of magnetite and $ J_{AB} $ is of a few meV.
We also note that the individual spin interaction energy 
$ J_{AB}S^{2} \cong 0.016 $ eV is comparable to the thermal energy 
at room temperature $ 0.026 $ eV. 
Thus, thermal effects are significant in the sintering process
for which temperatures are much above 300 K.

   To obtain a thermally equilibrated state, we minimize 
the internal energy of the spin system Eq.(1) at a given temperature 
using the Monte Carlo method with the Metropolis criterion. 
A trial random rotation is exerted on one of the spins 
in the n-th step: the trial is accepted if the internal energy 
decreases $ \delta U= U^{n} - U^{n-1} < 0$, or if the energy 
increase satisfies 
$ \exp ( - \delta U/k_{B}T) > \varepsilon, $
where $\varepsilon $ is a random number uniformly generated in 
the (0,1) interval, $ k_{B} $ is the Boltzmann constant and $ T $
is temperature; otherwise the trial is rejected. 

To proceed further, we have to specify the crystal structure and 
the exchange interaction coefficients. 
For magnetite, the unit cell is a cubic box with the sides of 
0.8396 nm at room temperature\cite{NIMS} and contains 24 irons: 
16 of them are ${\rm Fe}^{3+}$ ($3d^{5}, S=5/2$) and occupy all 
the tetrahedral sites (A-site, 8 positions) and half of the 
octahedral sites (B-site). 
The rest of irons are 8 ${\rm Fe}^{2+}$ ($3d^{6}, S=2$), and 
are located at B-sites. 
The exchange interaction coefficients are all negative and satisfy 
$ |J_{AB}| > |J_{AA}|, |J_{BB}| $\cite{Kittel}. 
We assume $ J_{AB}= -4.0 $ meV and $ J_{AA}= J_{BB}= -0.3 $ meV, 
where the former is larger in magnitude than the theoretically 
estimated value $ 2.5 $ meV\cite{Mills} to reproduce $ T_{c} $
in our model.

\section{Results}

   Using the procedures described in Sec.II, we first calculate 
the equilibrated state without an applied magnetic field for the 
periodic crystal of magnetite with $ 3 \times 3 \times 3 $ 
unit cells. 
Below $ T_{c} $, spins are ordered along the c-axis 
due to the exchange interactions, with the spins at A-sites and
those at B-sites oriented oppositely. 
The temperature dependence of magnetization calculated with our 
model is shown in Fig.1. 
The validity of our model is confirmed by the numerical result 
that the spontaneous magnetization decreases monotonically with 
temperature and vanishes above $ T_{c} $, as expected. 
Our calculated value $ M_{c}/M_{0} \cong 1.1 $ at 300 K roughly 
agrees with the experimental value $ M_{s}/M_{0} \cong 
1.3 $\cite{Kittel}, 
where $ M_{0}= N \mu_{B} $ with $ N $ the number of irons in 
the system.

\begin{figure}
\centerline{\scalebox{0.6}{\includegraphics{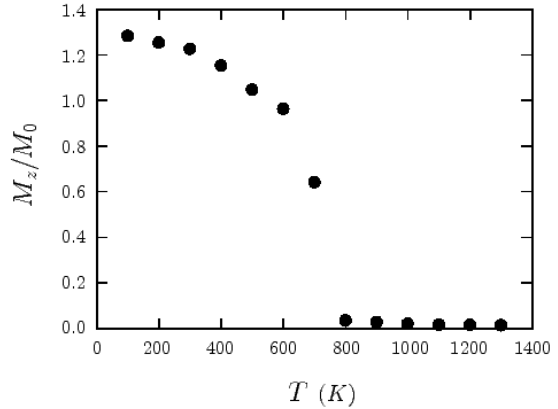}}}
\caption{
Temperature dependence of the magnetization along the 
c-axis (z-axis), $ M_{z}= - \sum_{i} g \mu_{B} s_{z,i} $, 
calculated by Monte Carlo simulation using the Heisenberg model 
for magnetite described 
in the text. The magnetization is normalized by $ M_{0}= N \mu_{B}$, 
where $ N $ is the number of Fe ions. The magnetization vanishes 
above the Curie temperature $ T_{c} $ of magnetite 858K. 
}
\end{figure}

\begin{figure}
\centerline{\scalebox{0.7}{\includegraphics{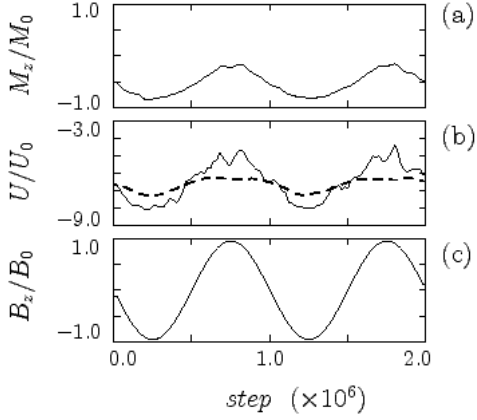}}}
\caption{
Model calculation for the magnetic field effects on 
magnetite at 700 K. The changes in (a) the z-component of 
magnetization $ M_{z}= - \sum_{i} g \mu_{B} s_{z,i} $, 
(b) the internal energy $ U $ per Fe (solid line), and the 
contribution of the Zeeman term to the internal energy 
(dashed line, for which the zero baseline is shifted downward 
by $ 7U_{0} $) are shown against (c) the applied magnetic field 
$ B_{z} $ in the Monte Carlo simulation steps. 
The z-axis is parallel to the c-axis. The normalization factors 
are as follows: $ M_{0}= N \mu_{B}, \
U_{0}= 10^{-21} $ J, and $ B_{0}= 1 $ T.
}
\end{figure}

   Next, we apply a slowly varying magnetic field $ {\bf B}_{w} $ 
to the equilibrium obtained above, and perform the Monte Carlo 
simulation. 
The varying speed of the magnetic field is chosen to be slow enough 
for spins to relax to the state of an instantaneous magnetic field.
The amplitude of the magnetic field is 1 T unless otherwise
specified, which is  still by two orders of magnitude less than 
the Weiss field. 
Spins relax in orientation under the magnetic field and 
the internal energy $ U({\bf B}_{w}) $ depends on the magnitude 
and orientation of $ {\bf B}_{w}$. 
We evaluate the difference between the maximum and minimum of 
the internal energy during a period of the magnetic field
change,
\begin{eqnarray}
\label{eq.3}
\Delta U^{(mc)} = U^{(mc)}_{max} - U^{(mc)}_{min}.
\end{eqnarray}
Here, the superscript (mc) stands for "Monte Carlo" simulation.
The energy $ \Delta U $ that is to be released irreversibly to
lattice atoms by dissipation process, possibly by spin-lattice
interactions\cite{Fivez}, is obtained by subtracting 
from $ \Delta U^{(mc)} $ the reversible energy change, 
$ \Delta U^{(rev)} = U^{(rev)}_{max} - U^{(rev)}_{min} $.
The latter is obtained by solving the spin dynamics equation
without dissipation,
\begin{eqnarray}
\label{eq.4}
   d{\bf s}_{i}/dt = \sum_{j} (2J_{ij}/ \hbar) 
   {\bf s}_{i} \times {\bf s}_{j}             
   - (g \mu_{B}/ \hbar) {\bf s}_{i} \times {\bf B}_{w}.
\end{eqnarray}
The energy difference $ \Delta U $ can be used as the index of 
heating by the microwave magnetic field.
This is verified later by agreement of the estimation based on 
the imaginary part of the magnetic susceptibility with
the $ \Delta U $ here and also the experimental value of 
the heating rate. 

Figure 2 shows the changes in (a) the average magnetization in 
the z direction (c-axis), (b) the total internal 
energy (solid line) and the contribution of the Zeeman 
term (dashed line) at 700 K against (c) the alternating 
magnetic field which is parallel to the c-axis.
In the present case, the magnetization $ M_{z} $ stays 
negative along the z-axis, and the Zeeman term 
$ - {\bf M} \cdot {\bf B}_{w} $ takes positive or negative 
value according to whether $ {\bf B}_{w} $ is parallel or 
anti-parallel to the c-axis. 
In the phase where the applied magnetic field becomes more 
negative, the magnitude of magnetization increases due to 
alignment of spins parallel to the c-axis, and the internal energy 
becomes minimal in this phase of the magnetic field. 
We note that the large change in the internal energy occurs 
through the exchange interactions because the energy associated 
with the Zeeman term is small and reversible, with the former 
by a factor of $ J_{AB}S/g \mu_{B}B_{w} \ (\gg 1) $ larger than
the latter, as depicted in Fig.2(b). 
In fact, the contribution of the Zeeman term is small and 
roughly the same as that obtained with the dissipationless 
spin dynamics Eq.(3).

\begin{figure}
\centerline{\scalebox{0.8}{\includegraphics{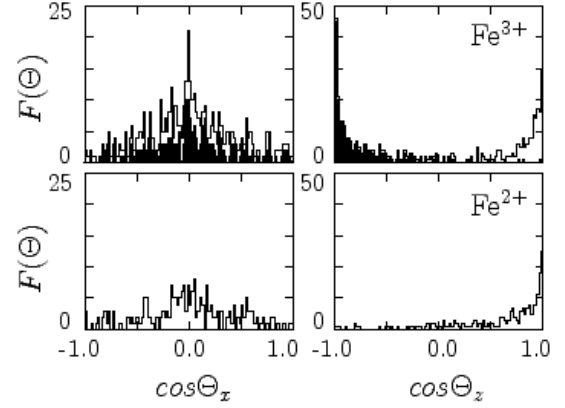}}}
\caption{
The distribution functions of spins along the x direction 
(leftward panels) and z direction (rightward panels), 
$ \Theta $ is the angle of the spin either with the x- or z-axis. 
The distribution functions in the upper panels correspond to 
$ {\rm Fe}^{3+} $ (the shaded area denotes the states of spins 
at A-sites) and those in the lower panels correspond to 
$ {\rm Fe}^{2+} $.
}
\end{figure}

\begin{figure}
\centerline{\scalebox{0.6}{\includegraphics{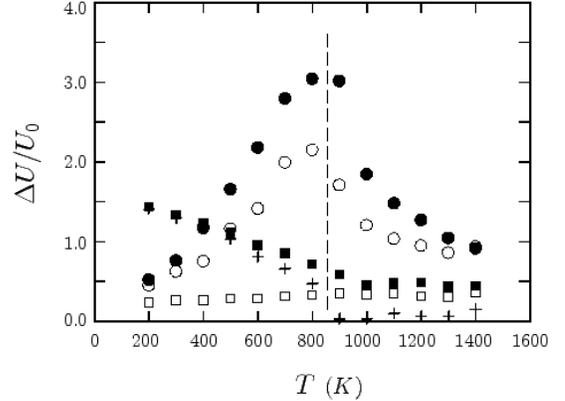}}}
\caption{
Temperature dependence of the internal energy change (per Fe)
during a period of an applied magnetic field of
$ B_{w}= 1 $ T ($ U_{0}= 10^{-21} $ J). 
The filled circles and squares correspond, respectively, to the
irreversible energy difference $ \Delta U $ and the change in
the Zeeman term $ \Delta({\bf M} \cdot {\bf B}_{w}) $ 
for magnetite with the magnetic field parallel to the c-axis; 
crosses correspond to $ \Delta U^{(rev)} $ obtained by  
the spin dynamics of Eq.(3) which is reversible.
The open circles and squares correspond to $ \Delta U $ and
the change in the Zeeman term, respectively, when the 
magnetic field is applied parallel to the a-axis.
The vertical line denotes $ T_{c} $ of magnetite. 
}
\end{figure}

\begin{figure}
\centerline{\scalebox{0.6}{\includegraphics{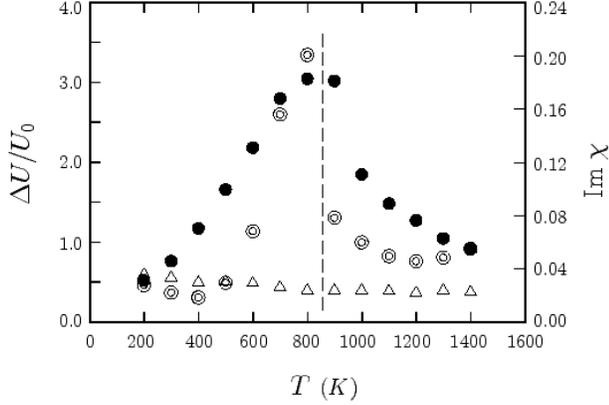}}}
\caption{
Temperature dependence of the internal energy change 
(per Fe) $ \Delta U $ for hematite is shown by triangles,
and that of magnetite with the magnetic field parallel 
to the c-axis is shown by filled circles as reference,
for $ B_{w}= 1 $ T ($ U_{0}= 10^{-21} $ J). 
The imaginary part of the magnetic susceptibility obtained 
by dissipative spin dynamics Eq.(4) for $ \tau_{D}= 1 $ ns 
is shown by double circles (in mol$^{-1}$). 
The vertical line denotes $ T_{c} $ of magnetite. 
}
\end{figure}

The distribution functions at the $ 1.25 \times 10^{6}$th 
Monte Carlo step in Fig.2 are shown in Fig.3. 
The x and z directions are taken along the a- and c-axes, 
respectively. 
In the x direction, the spins form centered Boltzmann distributions 
and no magnetization occurs along this direction. 
(The shaded areas correspond to the $ {\rm Fe}^{3+}$ at A-sites.) 
Ferrimagnetization occurs along the z direction. 
The $ {\rm Fe}^{3+} $ spins at A-sites are anti-parallel to 
the z-axis and those of $ {\rm Fe}^{2+}$ and $ {\rm Fe}^{3+} $ 
at B-sites are parallel to it. 
The change in the orientations in response to the microwave 
magnetic field is subtle, yet it gives rise to a maximum in the 
magnitude of magnetization periodically, as shown in 
Fig.2(a). 

Figure 4 shows the calculated temperature dependence 
of the internal energy difference $ \Delta U = 
\Delta U^{(mc)} - \Delta U^{(rev)} $ when the microwave 
magnetic field is parallel either to the c-axis 
(filled circles) or to the a-axis of magnetite 
(open circles). 
The difference in the energy becomes largest when the 
polarization of the magnetic field is parallel to the c-axis, 
and it increases linearly with temperature up to $ T_{c} $. 
The change in the reversible energy $ \Delta U^{(rev)} $ and 
that in the Zeeman term for the former case are shown 
by crosses and filled squares, respectively.
The energy change $ \Delta U^{(rev)} $ is similar to that 
in the spontaneous magnetization which decreases with 
temperature and vanishes above $ T_{c} $. 
The change in the Zeeman term is almost the same as
$ \Delta U^{(rev)} $ but is finite for $ T > T_{c} $. 
The Zeeman term in the Monte Carlo simulation remains finite
because an induced magnetization appears synchronously 
with and along the magnetic field independently of 
its polarization in the paramagnetic regime. 
The major contribution to the internal energy difference
$ \Delta U $ is attributed to the exchange interactions
since the change in the Zeeman term is reversible and thus 
almost subtracted in $ \Delta U $.
The temperature dependences shown in Fig.4 are in excellent 
agreement with the sintering experiments in the microwave 
magnetic field\cite{Agrawal1,Sato}, 
where the heating of magnetite was enhanced at 
$ 300-600^{\circ}C $ and continued to much above $ T_{c} $. 

\begin{figure}
\centerline{\scalebox{0.6}{\includegraphics{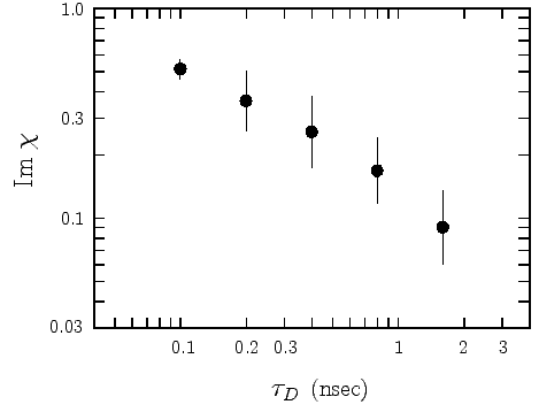}}}
\caption{
The relaxation time $ \tau_{D} $ dependence of the imaginary
part of magnetic susceptibility $ \chi $ (in mol$^{-1}$) 
for the 2.5GHz microwave and temperature 700K.
}
\end{figure}

\begin{figure}
\centerline{\scalebox{0.6}{\includegraphics{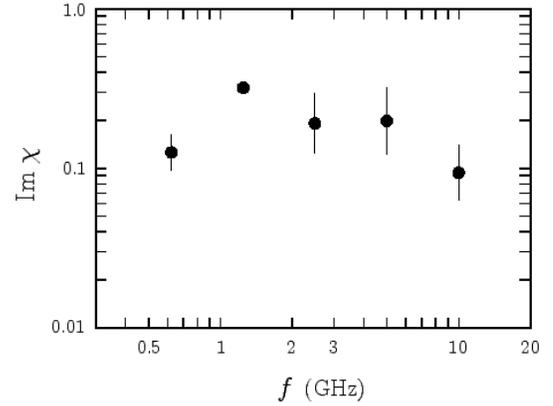}}}
\caption{
The microwave frequency $ \omega $ dependence of the imaginary
part of magnetic susceptibility $ \chi $ (in mol$^{-1}$) 
for the relaxation time $ \tau_{D} = 1$ ns and temperature 700K.
}
\end{figure}

An argument that involves time scales as well as the 
temperature dependence of heating is made possible 
by solving the dissipative spin dynamics. 
Here, the term $ -(\bf{s}_{i} -\bf{s}_{i0})/ \tau_{D} $ is
added to the righthand side of Eq.(3).
Namely, we solve 
\begin{eqnarray}
\label{eq.5}
   d{\bf s}_{i}/dt = \sum_{j} (2J_{ij}/ \hbar) 
   {\bf s}_{i} \times {\bf s}_{j}             
   - (g \mu_{B}/ \hbar) {\bf s}_{i} \times {\bf B}_{w} 
  \nonumber \\[-0.2cm]
   -(\bf{s}_{i} -\bf{s}_{i0})/ \tau_{D}.
\end{eqnarray}
A set of $ \bf{s}_{i0} $ constitutes the equilibrium 
spin distribution function for a given temperature and magnetic 
field under the Monte Carlo calculation.
The imaginary part of the magnetic susceptibility $ \chi_{i} $ 
is obtained from the linear response of magnetization
against the applied alternating magnetic field. 
$ \chi_{i} $ gives the heating rate $ dT/dt \propto \chi_{i} $,
and is shown for magnetite with double circles in Fig.5.
It peaks around 800 K and is consistent with 
$ \Delta U $ obtained with the energy principle of the
Monte Carlo calculation. 

The dependence of the imaginary part of magnetic susceptibility 
$ \chi_{i} $ on the relaxation time $ \tau_{D} $ is shown 
in Fig.6 for the 2.5 GHz microwave. 
We see that $ \chi_{i} $ is inversely proportional to
$ \tau_{D} $ in the $ \tau = 2\pi/\omega \le \tau_{D} $ range.  
Also, the dependence of $ \chi_{i} $ on the microwave frequency
is shown in Fig.7 for $ \tau_{D} = 1 $ ns.
The imaginary part of magnetic susceptibility peaks around
2 GHz.
This agrees with the experimentally deduced magnetic 
permeability for magnetite\cite{Ma}.
In our calculation, the relaxation time is assumed to be 
constant irrespectively of temperature or microwave frequency.
However, if the relaxation time becomes small with 
temperature rise, the heating rate increases as 
$ \tau_{D}^{-1} $ as shown in Fig.5, resulting in more 
rapid heating at elevated temperatures.

From these data, one obtains the heating rate
$ dT/dt \sim (1/2) \omega \chi_{i} B_{mw}^{2}/c_{p} \cong
300 $ K/s for the case of $ \tau_{D}= 1 $ ns and the
microwave field $ B_{mw}= $ 150 G,
where the heat capacity $ c_{p} $ is 210 J/K mol at 600K
(the wave period is 400 ps for the 2.45GHz microwave).
This is large enough to account for the experimental
value $ (dT/dt)_{ex} \cong 250 $ K/s in the microwave 
sintering of magnetite\cite{Agrawal1}.

\begin{figure}
\centerline{\scalebox{0.6}{\includegraphics{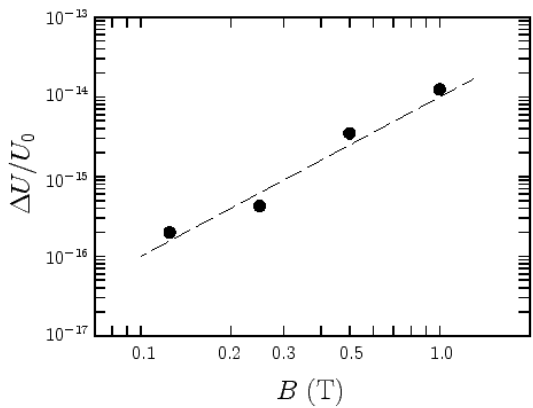}}}
\caption{
Magnetic field dependence of the internal energy change in
magnetite per wave period for the 2.5GHz microwave and
temperature 700 K.
The dotted line corresponds to $ B^{2} $.
}
\end{figure}

The dependence of the internal energy change in magnetite 
per wave period on the applied magnetic field is shown in Fig.8. 
The magnetic field amplitude ranges from 0.125 T to 1.0 T.
It is well fitted by the quadratic law $ B^{2} $, 
therefore, $ \Delta U $ thus the heating rate is proportional to 
the microwave power.

Hematite has a different crystal structure from magnetite\cite{NIMS}
and has weak spontaneous magnetization.
The calculated change in the internal energy $ \Delta U $  
is plotted for various temperatures by triangles in Fig. 5. 
By comparison of $ \Delta U $ with the case of magnetite, 
the response of hematite to the alternating magnetic field
is considerably weak. 
This is consistent with the experimental fact that only the domain 
of magnetite in a magnetite-hematite composite powder was heated 
by the magnetic field of microwaves\cite{Sato}. 

The observation that heating of titanium oxide occurs 
only when oxygen defects are present\cite{Agrawal2}, 
TiO$_{2-x}$ ($ x >0 $), is explained in a similar fashion.
Titanium in titanium oxide TiO$_{2} $ is a Ti$^{4+}$ ion
with $3d^{0}$ electron configuration, which has no electrons 
in the 3d shell. 
Thus, it has no spontaneous magnetization and should not respond 
to microwave magnetic field, similarly with hematite. 
However, when there are oxygen defects, the trivalent titanium 
ions $ {\rm Ti}^{3+} $ appear which have $3d^{1}$ electron
configuration.
The spins of these 3d electrons respond to the magnetic field 
of microwaves, and absorb microwave energy which leads to 
observed heating under the microwave magnetic field.

\section{Summary}

In this paper, we have showed theoretically the mechanism of
the rapid and selective sintering of magnetic metal oxide 
particles by the magnetic field of microwaves.
We adopt the Heisenberg model, and perform both Monte Carlo
calculation and dissipative spin dynamics simulations.
The heating occurs due to the response of magnetization 
to microwaves, which originates from electron spins residing 
in the unfilled 3d shell.
Their non-resonant response causes a large change in the 
internal energy through the exchange interaction between spins.
It persists above the Curie temperature $ T_{c} $ because 
each electron spin is able to respond to the alternating 
magnetic field of microwaves even above $ T_{c} $.
This energy change will then be dissipated to lattices and
contribute to heating.

Hematite $ {\rm Fe}_{2}{\rm O}_{3}$ which has only weak 
spontaneous magnetization shows much less response to 
microwaves than magnetite.
Also, the heating of titanium oxide having oxygen defects
TiO$_{2-x} $ (x>0) by the microwave magnetic field is 
explained by our theory in terms of intrinsic (spontaneous)
magnetization.

The imaginary part of the magnetic susceptibility Im$ \chi $ 
obtained by solving dissipative spin dynamics agrees well 
with the heating results by the Monte Carlo calculation. 
We have also presented the dependences of the heating rate 
on the frequency of microwaves and on the spin relaxation time.
These results well account for the large heating rate of
magnetic metal oxide by the microwave magnetic field 
in the sintering experiments.

\acknowledgments{
One of the authors (M.T.) is grateful to Prof.M.Sato,
Prof. I.Ohmine, Dr. M.Yamashiro and Dr. M.Ignatenko
for fruitful discussions.  
This work was supported by Grant-in-Aid for Prime Area 
Research No.18070005 from the Japan Ministry of Education, 
Science and Culture. 
}

\end{document}